# Towards a Dynamical Solution of the Strong CP Problem *

G. Schierholz

Deutsches Elektronen-Synchrotron DESY, D–22603 Hamburg, Germany
and
Gruppe Theorie der Elementarteilchen, Höchstleistungsrechenzentrum HLRZ,
c/o Forschungszentrum Jülich, D–52425 Jülich, Germany

It is argued that QCD might solve the strong CP problem on its own. To test this idea, a lattice simulation suggests itself. In view of the difficulty of such a calculation we have, as a first step, investigated the problem in the $CP^3$ model. The $CP^3$ model is in many respects similar to QCD. In this talk I shall present some first results of our calculation. Among other things it is shown that the model has a first order deconfining phase transition in $\theta$ and that the critical value of $\theta$ decreases towards zero as $\beta$ is taken to infinity. This suggests that $\theta$ is tuned to zero in the continuum limit.

## 1. INTRODUCTION

The discovery of instantons in non-abelian gauge theories [1] has led to the solution of the $U_A(1)$ problem in QCD [2]. Yet it created another puzzle: the strong CP problem.

Instantons represent tunneling events between vacua of different winding number $n$, which we denote by $|n\rangle$. They cause the perturbative vacuum $|0\rangle$ to become quantum mechanically unstable. The proper vacuum states, which are stable under any gauge invariant operation, are

$$|\theta\rangle = \sum_n \exp(i\theta n)|n\rangle. \qquad (1)$$

These so-called $\theta$ vacua correspond to the action

$$S_\theta = S - i\theta Q, \qquad (2)$$

---
*Talk given at the *Third KEK Topical Conference on CP Violation*, 16-18 November 1993, KEK, Tsukuba, Japan.

where $S$ is the standard action and

$$Q = -\frac{1}{32\pi^2}\int d^4x\,\epsilon_{\mu\nu\rho\sigma}\mathrm{Tr}F_{\mu\nu}F_{\rho\sigma} \in \mathbb{Z} \qquad (3)$$

is the topological charge. By convention I will choose $\theta$ to lie in the interval $[0,2\pi)$. The additional term in (2) breaks parity, time-reversal invariance and CP. In the presence of quarks the $\theta$ angle can be rotated into a phase of the quark mass matrix $M$ and vice versa, due to the $U_A(1)$ anomaly. The effective angle is in this case $\bar\theta = \theta + \arg\det M$. A priori $\bar\theta$ is a free parameter. Experimentally one obtains the upper bound $\bar\theta \leq 10^{-9}$ [3].

Why is $\bar\theta$ so small? Three possible explanations have been offered, all of which involve imposing global symmetries on the low energy physics. One explanation is that the $u$ quark is massless. In this case the action is invariant under a chiral transformation, and so the $\theta$ parameter can be rotated away. However, I do not think that a theory with a massless charged particle does exist [4]. In a second explanation [5] an additional U(1) symmetry is introduced which dynamically tunes $\bar{\theta}$ to zero. This implies the existence of a light particle, the axion [6]. If it exists, its mass must lie between $10^{-6}$ and $10^{-3}$ eV [7]. The third explanation is based on introducing a discrete CP symmetry which is broken spontaneously. The parameter $\bar{\theta}$ is then calculable and can be made small enough to be consistent with experiment [8].

In addition, several authors have sought a solution of the strong CP problem *within* QCD, that means without the introduction of new symmetries and particles [9]. It is almost certain that the QCD vacuum will undergo fundamental changes when $\theta$ is taken different from zero [10]. Some time ago 't Hooft and Mandelstam have conjectured [11] that the confinement phenomenon can be understood in terms of a color magnetic superconductor, in which color magnetic monopoles condense and color electric charges, i.e. quarks and gluons, are confined by a dual Meissner effect. More recently this picture of the QCD vacuum has been successfully tested by lattice simulations [12]. Not only does one find evidence for monopole condensation, but it could also be shown that the monopole currents around a flux tube satisfy a dual London equation [13]. In the $\theta$ vacuum these monopoles acquire a color electric charge of the magnitude $\theta/2\pi$ [14]. For $\theta \neq 0$ one would then expect [15] that the long-range color electric forces are screened by monopoles and that confinement is lost. Very likely this will cause the breakdown of asymptotic freedom as well.

Probably this is not the only way of understanding the QCD vacuum. For example, it might also be possible to describe the infra-red properties of the QCD vacuum in terms of instantons. But I would expect that one will find the same dependence on $\theta$. Indeed, in a model of the QCD vacuum based on semi-classical ideas it has been argued that in the $\theta$ vacuum the long-range color fields are screened by instantons [16], thus leading to similar conclusions. (A further example is given by the $CP^3$ model which I will discuss in detail below. This model possess instantons but no monopoles, and the vacuum can be understood as an instanton liquid.)

How would this solve the strong CP problem? If it proves to be true that confinement is lost for $\theta \neq 0$, then $\theta = 0$ is the only choice. The problem then is to understand why the confinement mode is chosen by nature. The answer could be that the theory admits a continuum limit only for $\theta = 0$.

To test this possibility, the natural way to proceed is to simulate the theory on the lattice. What makes this calculation very difficult, however, is the fact that the action is complex for non-vanishing values of $\theta$. As a result, standard lattice techniques are not immediately applicable. This has led us to investigate the problem in a simpler model first.

## 2. THE $CP^{N-1}$ MODEL: A 2-D IMAGE OF QCD

A model, which in many respects is similar to QCD, is the $CP^{N-1}$ model in two space-time dimensions. The common properties include (i) the existence of instantons and a vacuum angle $\theta$, (ii) asymptotic freedom, (iii) a dynamically generated mass gap, (iv) dimensional transmutation and (v) confinement. A linear potential is of course not difficult to achieve in two dimensions, but in the $CP^{N-1}$ model confinement arises without the presence of any fundamental gauge fields which is remarkable.

The $CP^{N-1}$ model describes N-component, complex scalar fields $z_a(x)$ of unit length:

$$\overline{z}_a(x) z_a(x) = 1, \ a = 1, \cdots, N. \qquad (4)$$

Out of these fields one may construct composite vector fields

$$A_\mu(x) = \frac{i}{2} \overline{z}_a(x) \stackrel{\leftrightarrow}{\partial}_\mu z_a(x). \qquad (5)$$

Then the action [17] can be cast into the intuitive

form [18]

$$S = \beta \int d^2x \overline{D_\mu z}_a(x) D_\mu z_a(x), \quad (6)$$

where $D_\mu = \partial_\mu + iA_\mu$. It follows that (6) is invariant under the gauge transformation $z'_a(x) = \exp(i\alpha(x)) z_a(x)$. Thus the model corresponds to a set of charged scalar fields interacting minimally with a composite gauge field. The topological charge is given by

$$Q = \frac{1}{2\pi} \int d^2x\, i\epsilon_{\mu\nu} \overline{D_\mu z}_a(x) D_\nu z_a(x) \quad (7)$$

$$= \frac{1}{2\pi} \int d^2x\, \epsilon_{\mu\nu} \partial_\mu A_\nu(x) \quad (8)$$

$$\equiv \frac{1}{2\pi} \int d^2x\, F_{01}. \quad (9)$$

This means that a field configuration with a non-trivial topological charge can also be seen as a configuration with a background electric field.

Confinement is directly related to the $\theta$ dependence of the theory, as we shall see in a moment. The $\theta$ dependence is essentially described by the partition function

$$Z(\theta) = \sum_Q \exp(i\theta Q)\, p(Q), \quad (10)$$

where

$$p(Q) = \frac{\int [\mathcal{D}z\mathcal{D}\overline{z}]_Q\, \delta(|z|^2 - 1)\, \exp(-S)}{\int \mathcal{D}z\mathcal{D}\overline{z}\, \delta(|z|^2 - 1)\, \exp(-S)}. \quad (11)$$

The subscript $Q$ means that the path integration is restricted to the charge sector indicated. The associated free energy per space-time volume $V$ is

$$F(\theta) = -\frac{1}{V} \ln Z(\theta). \quad (12)$$

The connected moments of the topological charge distribution are obtained from $F(\theta)$:

$$\frac{1}{V} \langle \theta | Q^n | \theta \rangle_c = -i^n \frac{d^n F(\theta)}{d\theta^n}. \quad (13)$$

In particular, the average charge density is given by

$$\frac{1}{V} \langle \theta | Q | \theta \rangle \equiv -iq(\theta) \quad (14)$$

$$= -i \frac{dF(\theta)}{d\theta}. \quad (15)$$

Consider now a pair of static external particles of charge $e$ and $-e$ in units of the intrinsic charge. The potential between the two particles is derived from the Wilson loop formula

$$W(A, \theta) = \langle \theta | \exp(ie \oint_{C(A)} dx_\mu A_\mu) | \theta \rangle, \quad (16)$$

where $A$ is the area enclosed by the contour $C(A)$. From (8) and (15) it then follows that

$$W(A, \theta) = \exp(-A \int_\theta^{\theta + 2\pi e} d\theta' \frac{dF(\theta')}{d\theta'}) \quad (17)$$

$$= \exp(-A[F(\theta + 2\pi e) - F(\theta)]) \quad (18)$$

which for $\theta = 0$ and small $e$ implies confinement by a linear potential. However, when $e = 1$, i.e. when $e$ is equal to the intrinsic charge, the long-range force is absent because $F(\theta)$ is periodic in $\theta$: $F(\theta) = F(\theta + 2\pi)$ [19]. The physical reason for that is that the external charges are screened by pair-produced $z$ particles. This is exactly the same situation as in QCD. The string tension is given by

$$\sigma(e, \theta) = F(\theta + 2\pi e) - F(\theta). \quad (19)$$

Note that the string tension is not invariant under charge conjugation, i.e. $\sigma(e, \theta) \neq \sigma(-e, \theta)$, for $\theta \neq 0, \pi$ due to the fact that the vacuum is an eigenstate of CP only for $\theta = 0, \pi$. Thus, if a particle of charge $e$ is to the left of an anti-particle of charge $-e$ it will feel a different force than if it is to the right of the anti-particle [20].

## 3. LATTICE SIMULATION WITH $\theta \neq 0$

The $CP^{N-1}$ model has been investigated by large-N [18] and instanton methods [21] in the past. In the former approach no spectacular $\theta$ dependence was observed in leading approximation. But this is not surprising. The higher moments ($n > 2$) of the topological charge distribution (13) are zero in this approximation which eliminates quantum effects to a great extent.

We have chosen to examine the $CP^3$ model in closer detail [22]. On the lattice the action takes the form

$$S = -2\beta \sum_{x,\mu} |\overline{z}_a(x) z_a(x+\hat{\mu})|^2 \qquad (20)$$

($a = 1, \cdots, 4$). The boundary conditions are taken to be periodic. In accordance with (5) one defines link fields

$$U_\mu(x) = \frac{\overline{z}_a(x) z_a(x+\hat{\mu})}{|\overline{z}_b(x) z_b(x+\hat{\mu})|} \qquad (21)$$

$$= \exp(\mathrm{i} A_\mu(x)). \qquad (22)$$

The topological charge is derived from (21) by forming parallel transporters around plaquettes,

$$U_\square = \exp(\mathrm{i} F_{01}), \quad -\pi < F_{01} \leq \pi, \qquad (23)$$

and writing

$$Q = \frac{1}{2\pi} \sum_\square F_{01}. \qquad (24)$$

It is easy to see that $Q$ is an integer. For this definition of $Q$ it is expected that the model does not give rise to dislocations [23]. Dislocations are lattice artifacts on the scale of the order of the lattice spacing which cause the free energy $F(\theta)$ to diverge in the continuum limit.

We compute $Z(\theta)$ by computing the probability function $p(Q)$. This is done by the following method [24]. We divide the phase space into overlapping sets of five consecutive charges. In each of these sets the $z$ fields are updated by a combination of Metropolis and overrelaxation steps, $z' = \exp(\mathrm{i} \phi_a \lambda_a) z$, where the $U(4)$ generators $\lambda_a$ are selected randomly [25]. This is supplemented by an additional acceptance criterion: if the new charge is in the same set the configuration is accepted and the new charge recorded; if the new charge is outside the set the change is rejected and the old charge recorded. Furthermore, a trial charge distribution which is approximately equal to the true distribution is incorporated.

The probability function $p(Q)$ is a steeply falling function of $Q$. With this method we were able to compute $p(Q)$ over a range of more than twenty orders of magnitude.

## 4. FIRST PROMISES

We are ready now for a quantitative test. If our idea is correct, we should find a first order phase transition in $\theta$ from a confining phase to a Higgs or Coulomb phase. On a finite lattice and at a finite value of $\beta$ the phase transition would occur at a value $\theta = \theta_c(\beta, V) > 0$, where $V$ now is the lattice volume. Only on an infinite lattice and at $\beta = \infty$, i.e. in the continuum limit, can we expect that $\theta_c = 0$.

A first order phase transition will manifest itself in a kink in the free energy, $F(\theta)$, as well as in a discontinuity in the first derivative of the free energy, $dF(\theta)/d\theta$. In Fig. 1 I show $F(\theta)$ on the $V = 64^2$ lattice at $\beta = 2.7$. [2] The correlation length for this value of $\beta$ is $\xi \approx 8.8$ [26]. For comparison I also show the prediction of the large-N expansion to leading order [18], which turns out to be $F(\theta) = c\,\theta^2$, where the constant $c$ has been fitted to the lattice data at small $\theta$. [3] We see a distinctly marked kink at $\theta = \theta_c \approx 0.5\,\pi$: while $F(\theta)$ increases roughly proportional to $\theta^2$ up to $\theta = \theta_c$, we find that $F(\theta)$ is constant (within the error bars) for $\theta > \theta_c$. The latter result is rather remarkable, as it means that all derivatives of $F(\theta)$ vanish for $\theta > \theta_c$. In other words, the theory becomes independent of $\theta$. The large-N expansion, on the other hand, predicts a phase transition at $\theta = \pi$. In Fig. 2 I show the first derivative of the free energy. According to (15) this quantity is identical to the topological charge density, $q(\theta)$, which again can be interpreted as a background electric field (cf. (9) and (24)). We see that $q(\theta)$ increases almost linearly with $\theta$ up to $\theta = \theta_c$, where it jumps to zero and then stays zero over the rest of the interval (again within the error bars). Thus the phase transition is accompanied by a collapse of the background electric field which hints at a transition from a confining phase to a Higgs phase.

---

[2]The reader who recalls Ref. [22] might notice a slight change in the shape of $F(\theta)$. The explanation is that in the meantime we have increased our statistics by an order of magnitude.
[3]This gives $c = (1/2)\,d^2 F(\theta)/d\theta^2|_{\theta=0} = (1/2)\,\langle 0|Q^2|0\rangle/V$.

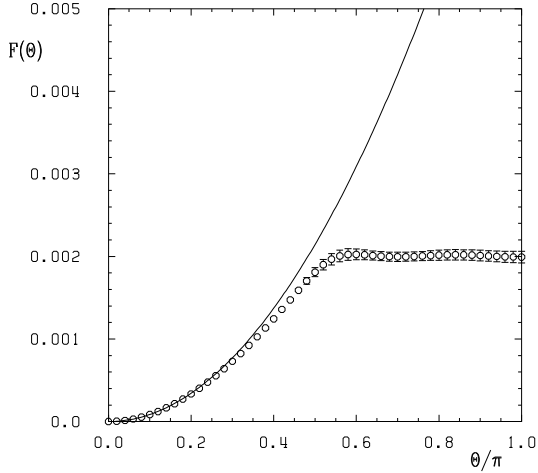
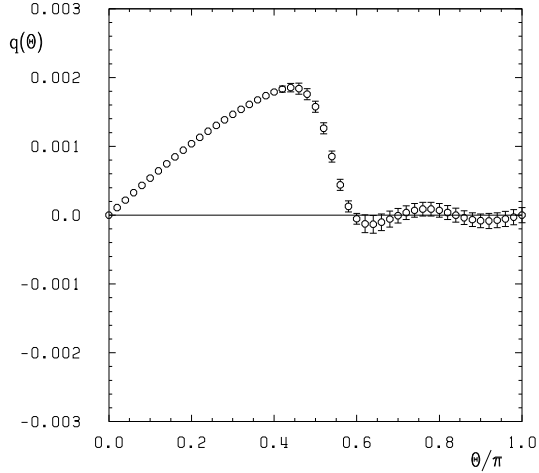

Figure 1. The free energy $F(\theta)$ as a function of $\theta$ on the $V = 64^2$ lattice at $\beta = 2.7$. The solid curve is the prediction of the large-N expansion to leading order. Only the first half of the $\theta$ interval is displayed. In the second half of the interval $F(\theta) = F(2\pi - \theta)$.

Figure 2. The topological charge density $q(\theta)$ as a function of $\theta$ on the $V = 64^2$ lattice at $\beta = 2.7$. Only the first half of the $\theta$ interval is displayed. In the second half of the interval $q(\theta) = -q(2\pi - \theta)$.

Before we can make any statements about the continuum, we have to extrapolate the lattice results to $V = \infty$ and $\beta = \infty$. For a first order phase transition we expect

$$\theta_c(\beta, V) - \theta_c(\beta, \infty) \propto V^{-1} \qquad (25)$$

for fixed $\beta$. In Fig. 3 I show the results for $\theta_c(\beta, V)$ for a variety of lattice volumes ranging from $V = 72^2$ to $V = 28^2$ and for two values of $\beta$, $\beta = 2.5$ and $\beta = 2.7$. For $\beta = 2.5$ the correlation length is $\xi \approx 4.5$ [26]. I have chosen to plot $\theta_c(\beta, V)$ as a function of $V^{-1}$. We see that for both values of $\beta$ our data fall on a straight line in accordance with a first order phase transition, thus confirming our earlier statement about the order of the transition. This allows us to extrapolate the lattice results to the infinite volume. We obtain $\theta_c/\pi = 0.32(2)$ for $\beta = 2.5$ and $\theta_c/\pi = 0.18(3)$ for $\beta = 2.7$. These values are to be contrasted with the analytic result in the strong coupling region $\theta_c = \pi$ [27].

Three values of $\beta$ are not enough to make a precise extrapolation to $\beta = \infty$. But the trend is evident: $\theta_c$ depends strongly on $\beta$, and $\theta_c$ decreases towards zero as we approach the continuum limit. Our resuls are consistent with a decay like $\theta_c \propto 1/\xi$. This suggests a phase diagram of the form sketched in Fig. 4. The horizontal line at $\theta = \pi$ is the prediction of the strong coupling calculation. It is a line of first order phase transitions on which CP invariance is spontaneously broken. Whether $\theta_c$ is indeed exactly zero at $\beta = \infty$ we cannot say at the moment. But our results so far are promising. In the near future it should be possible to repeat the calculation for a correlation length that is up to a factor of five times larger than that at $\beta = 2.7$ [25, 26].

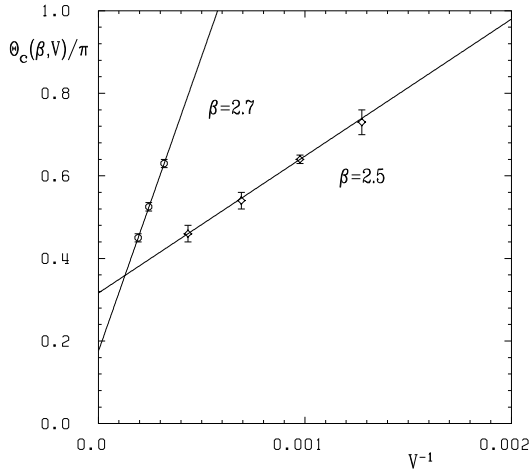
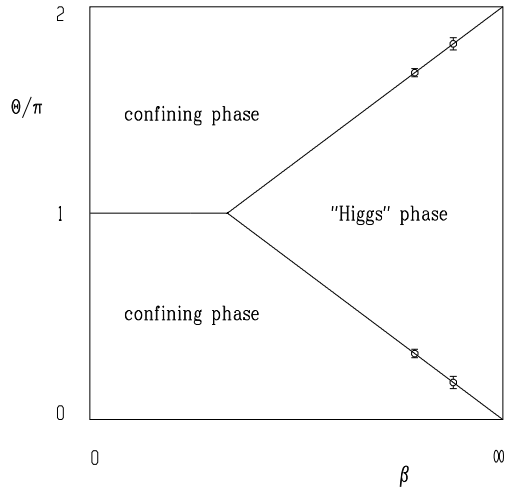

Figure 3. The critical value $\theta_c(\beta,V)$ as a function of $V^{-1}$ for two values of $\beta$. The lines are a linear fit to the data points.

Figure 4. The phase diagram over the full interval $0 \leq \theta < 2\pi$. The horizontal line at $\theta = \pi$ is the strong coupling prediction, and the symbols are the lattice data extrapolated to infinite volume.

To show that the phase transition is indeed a deconfining phase transition, let me discuss the string tension now. In (19) I have derived a formula that relates the string tension to the free energy. According to this formula it is sufficient to know $F(\theta)$ over a wide enough range of $\theta$. This we do already. To keep the discussion simple, I shall restrict myself to the case of small fractional charge $e$. This matters because the vacuum will change its properties if it is exposed to a strong external electric field, as we have seen before. The $\theta$ dependence of the string tension can then be read off from Fig. 1. From that figure and (19) it follows that at $\theta = 0$ the string tension is $\sigma(e,0) \approx c(2\pi e)^2 = 2\pi^2 e^2 \chi_t$, where $\chi_t = \langle 0|Q^2|0\rangle/V$ is the topological susceptibility.

This relation becomes exact in the limit $e \to 0$.[4] If we now increase $\theta$, the string tension will increase until $\theta$ reaches $\theta_c - 2\pi e$, where it will start to decrease again. Finally, at $\theta = \theta_c$ the string tension will be zero. And provided the external charge is small enough, it will remain zero up to $\theta = \pi$. This result is a consequence of the property that $F(\theta)$ is constant for $\theta > \theta_c$. Let us now look at the data for the string tension directly. In Fig. 5 I show $\sigma(e,\theta)$ for two charges, $e = 0.1$ and $e = 0.2$, on the $V = 64^2$ lattice at $\beta = 2.7$. Remember that on this lattice $\theta_c$ was $\approx 0.5\,\pi$. We clearly see that the string tension vanishes for $\theta > \theta_c$ (within the error bars).

We have also investigated the $\beta$ function and its dependence on $\theta$. We find evidence that the $\beta$ function vanishes at the phase transition, indicating that asymptotic freedom might also be lost in the so-called Higgs phase. Because of lack of space I cannot go into detail here. A full account of our work will be given elsewhere [28].

---

[4]Note that this is precisely the result of the large-N expansion to leading order [18].

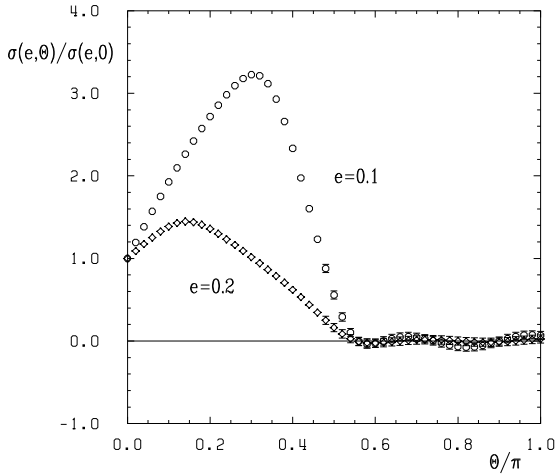

Figure 5. The string tension $\sigma(e,\theta)$ as a function of $\theta$ for $e = 0.1$ and $e = 0.2$ on the $V = 64^2$ lattice at $\beta = 2.7$. Only the first half of the $\theta$ interval is displayed. In the second half of the interval $\sigma(e,\theta) = \sigma(-e, 2\pi - \theta)$.

## 5. SUMMARY AND OUTLOOK

Let me now summarize our results. We have found a phase transition in $\theta$ from a confining phase to a non-confining phase, presumably a Higgs phase. This by itself is already a remarkable result which has interesting applications. The critical value of $\theta$ turns out to vary strongly with $\beta$. While $\theta_c = \pi$ in the strong coupling region, $\theta_c$ falls off rapidly in the direction of zero as $\beta$ is taken to infinity. This behavior is summarized in the phase diagram shown in Fig. 4. The important question that remains to be answered is whether the line of first order phase transitions will indeed end in the point $(\beta, \theta) = (\infty, 0)$.[5] If this proves to be the case, then $\theta = 0$ is the only point at which the continuum limit can be taken, at least in the confining phase. This would resolve the strong CP problem. As far as other possible fixed points are concerned, there are indications that the theory is not asymptotically free in the Higgs phase. I would therefore not expect to find any other ultra-violet stable fixed point.

So far we have focussed all our efforts on the $CP^3$ model. The algorithm we have employed has proven to be very efficient. I am confident that one will be able to solve the strong CP problem one day also in QCD, though the demand in computing power will be vast. As a first step in that direction we are currently investigating the problem in the SU(2) Yang-Mills theory.

## ACKNOWLEDGEMENTS

The numerical calculations in the $CP^3$ model were done together with Štefan Olejník whom I wish to thank for a fruitful collaboration. The computer time was granted to us by HLRZ.

---

[5] I consider only the lower branch of the phase transition line here.

## REFERENCES

1  A. Belavin, A. Polyakov, A. Schwartz and Y. Tyupkin, Phys. Lett. **B59** (1975) 85.
2  G. 't Hooft, Phys. Rev. Lett. **37** (1976) 8; Phys. Rev. **D14** (1976) 3432.
3  V. Baluni, Phys. Rev. **D19** (1979) 2227; R. Crewther, P. Di Vecchia, G. Veneziano and E. Witten, Phys. Lett. **B88** (1979) 123; Erratum, *ibid.* **B91** (1980) 487.
4  M. Göckeler, R. Horsley, P. Rakow, G. Schierholz and R. Sommer, Nucl. Phys. **B371** (1992) 713; A. Ali Khan, Thesis (Hamburg, 1994)
5  R. Peccei and H. Quinn, Phys. Rev. Lett. **38** (1977) 1440.
6  S. Weinberg, Phys. Rev. Lett. **40** (1978) 223; F. Wilczek, Phys. Rev. Lett. **40** (1978) 279.
7  E.g., M. Turner, Phys. Rep. **C197** (1990) 1.
8  A. E. Nelson, Phys. Lett. **B163** (1984) 387; S. M. Barr, Phys. Rev. Lett. **53** (1984) 329; K. Babu and R. N. Mohapatra, Phys. Rev. **D41** (1990) 1286; S. M. Barr, D. Chang and G. Senjanovic, Phys. Rev. Lett. **67** (1991) 2765.

9.  Y.-S. Wu and A. Zee, Nucl. Phys. **B258** (1985) 157;
    S. Samuel, Mod. Phys. Lett. **A7** (1992) 2007.
10. G. 't Hooft, Nucl. Phys. **B190** (1981) 455.
11. G. 't Hooft, in *High Energy Physics*, Proceedings of the EPS International Conference on High Energy Physics, Palermo, 1975, ed. A. Zichichi (Editrice Compositori, Bologna, 1976);
    G. 't Hooft, Phys. Scripta **25** (1982) 133;
    S. Mandelstam, Phys. Rep. **C23** (1976) 245.
12. A. S. Kronfeld, G. Schierholz and U.-J. Wiese, Nucl. Phys. **B293** (1987) 461;
    A. S. Kronfeld, M. L. Laursen, G. Schierholz and U.-J. Wiese, Phys. Lett. **B198** (1987) 516;
    F. Brandstaeter, G. Schierholz and U.-J. Wiese, Phys. Lett. **B272** (1991) 319;
    for a recent review see:
    T. Suzuki, Nucl. Phys. **B** (Proc. Suppl.) **30** (1993) 176, and references therein.
13. V. Singh, D. A. Browne and R. W. Haymaker, Phys. Lett. **B306** (1993) 115.
14. E. Witten, Phys. Lett. **B86** (1979) 283.
15. Z. F. Ezawa and A. Iwazaki, Phys. Rev. **D26** (1982) 631.
16. C. G. Callan, R. F. Dashen and D. J. Gross, Phys. Rev. **D20** (1979) 3279.
17. H. Eichenherr, Nucl. Phys **B146** (1978) 215;
    E. Cremmer and Scherk, Phys. Lett. **B74** (1978) 341.
18. A. D'Adda, P. Di Vecchia and M. Lüscher, Nucl. Phys. **B146** (1978) 63;
    E. Witten, Nucl. Phys. **B149** (1979) 285.
19. M. Lüscher, Phys. Lett. **B78** (1978) 465.
20. S. Coleman, Ann. Phys. **101** (1976) 239.
21. B. Berg and M. Lüscher, Commun. Math. Phys. **69** (1979) 57;
    I. Affleck, Nucl. Phys. **B162** (1980) 461.
22. Š. Olejník and G. Schierholz, DESY preprint 93-195 (1993); to be published in Nucl. Phys. **B** (Proc. Suppl.).
23. D. Petcher and M. Lüscher, Nucl. Phys. **B225** (1983) 53.
24. G. Bhanot, S. Black, P. Carter and R. Salvador, Phys. Lett. **B183** (1987) 331;
    G. Bhanot, K. Bitar and R. Salvador, Phys. Lett. **B187** (1987) 381; *ibid.* **B188** (1987) 246;
    M. Karliner, S. Sharpe and Y. F. Chang, Nucl. Phys. **B302** (1988) 204;
    U.-J. Wiese, Nucl. Phys. **B318** (1989) 153.
25. M. Hasenbusch and S. Meyer, Phys. Rev. Lett. **68** (1992) 435.
26. U. Wolff, Phys. Lett. **B284** (1992) 94.
27. N. Seiberg, Phys. Rev. Lett. **53** (1984) 637.
28. Š. Olejník and G. Schierholz, in preparation.